\documentstyle[12pt]{article}
\begin{document}

\newcommand\beq{\begin{equation}}
\newcommand\eeq{\end{equation}}
\newcommand\bea{\begin{eqnarray}}
\newcommand\eea{\end{eqnarray}}
\newcommand\non{\nonumber}
\newcommand\bib{\bibitem}
\newcommand\df{$\delta$-function }
\newcommand\dpf{$\delta^{\prime \prime}$-function }

\begin{center}
{\bf The fermionic limit of the \df Bose gas: a pseudopotential approach}
\end{center}

\vskip 1 true cm
\centerline{Diptiman Sen \footnote{E-mail address: diptiman@cts.iisc.ernet.in}}

\vskip .5 true cm
\centerline{\it Centre for Theoretical Studies,}
\centerline{\it Indian Institute of Science, Bangalore 560012, India} 

\vskip 1 true cm
\begin{abstract}

We use first-order perturbation theory near the fermionic limit of the \df Bose
gas in one dimension (i.e., a system of weakly interacting fermions) to study 
three situations of physical interest. The calculation is done using a 
pseudopotential which takes the form of a two-body \dpf interaction. The three
cases considered are the behavior of the system with a hard wall, with a point
where the strength of the pseudopotential changes discontinuously, and with a 
region of finite length where the pseudopotential strength is non-zero (this
is sometimes used as a model for a quantum wire). In all cases, we obtain exact
expressions for the density to first order in the pseudopotential strength. The
asymptotic behaviors of the densities are in agreement with the results 
obtained from bosonization for a Tomonaga-Luttinger liquid, namely, an 
interaction dependent power-law decay of the density far from the hard wall, a
reflection from the point of discontinuity, and transmission resonances for the
interacting region of finite length. Our results provide a non-trivial 
verification of the Tomonaga-Luttinger liquid description of the \df Bose gas 
near the fermionic limit.

\end{abstract}
\vskip .5 true cm


\newpage

\vskip .5 true cm
\noindent{\bf 1. Introduction}
\vskip .3 true cm

The \df Bose gas in one dimension has been studied extensively ever since Lieb
and Liniger solved it using the Bethe ansatz \cite{lieb}. However, the wave 
functions which one obtains from the Bethe ansatz are usually quite difficult 
to work with. It is therefore useful to examine other ways of studying the \df
Bose gas. With this in mind, we recently developed a way of doing perturbation
theory near the fermionic limit of the model \cite{sen}, which is equivalent to
a Fermi gas with weakly attractive interactions. This involves the use of a 
two-body pseudopotential which takes the form of a \dpf; let us denote the 
strength of the pseudopotential by a parameter $g$ which will be defined below.
In Ref. 2, we showed that this perturbative approach correctly reproduces the
ground state energy up to order $g^2$. In this paper, we will apply the 
pseudopotential approach to three situations which cannot be solved using the
Bethe ansatz; this will illustrate the power of this approach. 

The \df Bose gas is an example of a Tomonaga-Luttinger liquid (TLL)
\cite{gogolin,hald}. A TLL is characterized by two quantities, the velocity of
the low-energy excitations $v$ (the dispersion relation of these excitations 
is given by $\omega = v |k|$), and the interaction parameter $K$. Once 
these two parameters are known, the low-energy and long wavelength properties 
of a TLL can be found by the technique of bosonization \cite{gogolin,hald}. 
In particular, $K$ determines the exponents governing the long distance 
behavior of various correlation functions. We will show that the results
obtained using the pseudopotential approach are in complete agreement with
those expected from bosonization. Our study will therefore provide a 
non-trivial check of the pseudopotential approach and will also confirm the 
expression for the parameter $K$.

The plan of this paper is as follows. In section 2, we will introduce the \df
Bose gas and the pseudopotential approach for doing perturbation theory near
the fermionic limit. In section 3, we will consider the \df Bose gas with a 
hard wall. We will obtain an expression for the density which is exact to first
order in $g$. We will then find the asymptotic behavior of the density far from
the hard wall. The expression of the density will involve logarithmic factors,
which, in section 6, will be recognized as being due to an interaction 
dependent power-law decay of the density which is characteristic of a TLL. In
section 4, we will consider a model in which the pseudopotential parameter $g$
changes discontinuously at one point. We will again compute the density exactly
to first order in $g$ and show that there are oscillations which can be 
interpreted as being due to a reflection of the particles from that point. The
amplitudes of reflection from the two sides are found to be equal to $\pm g$.
In section 5, we will extend the model of section 4 to the case of a
region of finite length over which the pseudopotential has a non-zero strength.
Finally, in section 6, we will discuss the TLL approach to the \df Bose gas, 
and will compute the Luttinger parameters $K$ and $v$ to first order in $g$. We
will then use bosonization to compute the asymptotic behaviors of the density 
far from a hard wall and from a point of discontinuity in the Luttinger 
parameters. We will then see that these results agree precisely with those 
obtained in the previous sections. We will conclude in section 7 by pointing
out some possible directions for future research.

\newpage

\noindent {\bf 2. The \df Bose gas}
\vskip .3 true cm

The \df Bose gas is defined by the Hamiltonian 
\beq
H ~=~ -~\frac{1}{2m} ~\sum_{1 \le i \le N} ~\frac{\partial^2}{\partial x_i^2}~
+~ \frac{2c}{m} ~\sum_{1 \le i < j \le N} ~\delta ~(~x_i ~-~ x_j ~) ~,
\label{ham}
\eeq
governing $N$ identical bosons moving in one dimension. (We have set $\hbar =
1$). The interaction parameter $c$ has the dimensions of (length)$^{-1}$. We 
will assume that $c$ is non-negative, otherwise the thermodynamic limit 
($N \rightarrow \infty$) is not well-defined due to the presence of bound 
states with arbitrarily large negative energies.

We will consider two different boundary conditions for the calculations 
presented in this paper. In one case, the particles will be considered to be on
a circle of circumference $L$ with the wave functions satisfying periodic 
boundary conditions. In the other case, the particles will be considered to be
in a box of length $L$ with the wave functions vanishing if any of the 
particles coordinates is equal to $0$ or $L$ (this is called the hard wall 
condition). Since the particles are identical bosons, the wave functions must 
be completely symmetric. There are $N!$ possible orderings of the particle 
coordinates, given by $0 \le x_{P_1} \le x_{P_2} \le \cdot \cdot \cdot \le 
x_{P_N} \le L$, where $(P_1,P_2, \cdot \cdot \cdot , P_N)$ is some permutation
of the numbers $(1,2, \cdot \cdot \cdot ,N)$. If the wave functions are known 
for any one ordering, say $0 \le x_1 \le x_2 \le \cdot \cdot \cdot x_N \le L$,
they are known for all other orderings by symmetry.

It is clear that the model describes noninteracting bosons for $c=0$. For $c=
\infty$, the wave functions vanish whenever any two particle coordinates 
coincide. We can then carry out the unitary transformation $\psi_P \rightarrow
(-1)^P \psi_P$, where $\psi_P$ denotes the wave function for the ordering $P$,
and $(-1)^P$ denotes the sign of the permutation $P$. Under this 
transformation, the wave function becomes completely antisymmetric, i.e., 
fermionic. Thus $c=\infty$ denotes a system of noninteracting fermions. (Note 
that this unitary transformation of the wave functions is only allowed if $c=
\infty$. At any other value of $c$, the symmetric wave functions do not vanish
for $x_i = x_j$, and the transformation would produce antisymmetric wave 
functions which are discontinuous at those coincident points). Henceforth, we 
will refer to the model with $c=\infty$ as being noninteracting, and the model
with $c$ large as being weakly interacting in the fermionic sense.

In Ref. 2, we developed a way of doing perturbation theory near the fermionic
limit in powers of $1/c$. We showed that the perturbation around $c=\infty$ is
described by a two-particle interaction of the form
\beq
V ~=~ -~ \frac{1}{mc} ~\sum_{1 \le i<j \le N} ~\delta^{\prime \prime} ( ~
x_i ~ -~ x_j ~)~.
\label{pp1}
\eeq
This pseudopotential can be used straightforwardly to first order in $1/c$. At
second and higher orders, some divergences appear which can be cured by a 
point-splitting prescription \cite{sen}. In this paper, we will work only to 
first order in $1/c$ and will therefore not encounter any divergences.

We will be interested in the thermodynamic limit in which $N, L \rightarrow 
\infty$ with the density 
\beq
\rho_0 ~=~ \frac{N}{L}
\eeq
held fixed. In the thermodynamic limit, we find, either from the exact 
solution \cite{lieb} or from the pseudopotential approach \cite{sen}, that 
the ground state energy per particle is given by
\beq
\frac{E_0}{N} ~=~ \frac{\pi^2 \rho_0^2}{6m} ~-~ \frac{\pi^2 \rho_0^3}{3mc}
\label{e0}
\eeq
to first order in $1/c$. The chemical potential $\mu = (\partial E_0 /\partial
N)_L$ is given by
\beq
\mu ~=~ \frac{\pi^2 \rho_0^2}{2m} ~-~ \frac{4\pi^2 \rho_0^3}{3mc} ~.
\label{mu}
\eeq
For convenience, we define the Fermi momentum $k_F$ and Fermi velocity $v_F$ as
\bea
k_F ~&=&~ \pi \rho_0 ~, \non \\
{\rm and} ~~~ v_F ~&=&~ \frac{k_F}{m} ~.
\eea
We also define a dimensionless parameter
\beq
g ~=~ \frac{\rho_0}{c} ~.
\eeq
Equations (\ref{e0}-\ref{mu}) imply that for large values of $c$, we have a 
system of fermions with a weak attractive interaction of strength $-g$.

The energy eigenstates of the system of noninteracting fermions ($g=0$) have
normalized wave functions which are given by $1/\sqrt{N!}$ times the 
Slater determinant of a matrix $M$. The entries of $M$ are given by
\beq
M_{np} ~=~ \psi_n (x_p) ~,
\eeq
where $\psi_n (x)$ denotes the normalized one-particle wave functions.
For the model on a circle, $\psi_n$ is given by 
\beq
\psi_n (x) ~=~ \frac{1}{\sqrt L} ~e^{i2\pi nx/L}
\label{wf1}
\eeq
where $n=0,\pm 1,\pm 2, \cdot \cdot \cdot$ (if $N$ is odd), and the 
corresponding one-particle energies are $E_n = (2\pi n)^2 /(2mL^2)$. For the 
system in a box with hard walls, the one-particle wave functions are given by
\beq
\psi_n (x) ~=~ {\sqrt \frac{2}{L}} ~\sin (\frac{\pi nx}{L})
\label{wf2}
\eeq
where $n=1,2,3, \cdot \cdot \cdot$, and the corresponding energies are 
$E_n = (\pi n)^2 /(2m L^2)$. 

Using the pseudopotential given in (\ref{pp1}), we can do perturbation theory 
to first order in $1/c$ as follows. For the noninteracting and weakly 
interacting systems, let us denote the $N$-particle wave functions of the 
energy eigenstates by $\Psi_n^{(0)}$ (which are normalized to 
unity) and $\Psi_n$ respectively, and the corresponding 
energies by $E_n^{(0)}$ and $E_n$. To first order in $1/c$, we have
\beq
\Psi_n ~=~ \Psi_n^{(0)} ~+~ \sum_{l \ne n} ~ \Psi_l^{(0)} ~\frac{\langle 
\Psi_l^{(0)} \vert V \vert \Psi_n^{(0)} \rangle}{E_n^{(0)} ~-~ E_l^{(0)}} ~,
\label{pert}
\eeq
provided that there is no degeneracy at the energy $E_n^{(0)}$. (This will be 
true in our calculations since we will only apply (\ref{pert}) to the ground 
state which is unique both on the circle and in the box). Note that the norm of
$\Psi_n$ in (\ref{pert}) differs from unity only by terms of order $1/c^2$ 
which we are going to ignore. 

The one-particle density $\rho(x)$ is obtained from the many-particle
wave function $\Psi (x_i)$ as 
\beq
\rho (x) ~=~ N ~\int ~dx_2 dx_3 \cdot \cdot \cdot dx_N ~\Psi^\star \Psi ~,
\label{dens1}
\eeq
where the factor of $N$ has been introduced on the right hand side to ensure
that we obtain $\int_0^L dx \rho (x) = N$. For the system on a circle, we find
that the density in the ground state of $N$ noninteracting fermions is simply 
equal to a constant, $\rho (x) = \rho_0$. For the system in a box,
the density in the ground state of $N$ noninteracting fermions is given by
\beq
\rho (x) ~=~ \sum_{n=1}^N ~\frac{2}{L} ~\sin^2 (\frac{\pi nx}{L}) ~.
\eeq
In the thermodynamic limit, we then find that 
\beq
\rho (x) ~=~ \rho_0 ~-~ \frac{\sin (2 k_F x)}{2\pi x} ~.
\label{dens2}
\eeq 
The hole in the density integrates to $1/4$, namely,
\beq
\int_0^\infty ~dx ~[~ \rho (x) ~-~ \rho_0 (x) ~]~ =~ - \frac{1}{4} ~.
\eeq
The oscillations in the density in (\ref{dens2}) are caused by reflection
from the hard wall at $x=0$. We will study below how these oscillations are
modified by the interactions between the particles.

\vskip .5 true cm
\noindent{\bf 3. The \df Bose gas with a hard wall}
\vskip .3 true cm

We will now study the behavior of the \df Bose gas with a hard wall. 
We will first consider the system placed in a box extending
from $x=0$ to $x=L$ with hard walls at both ends. We will then take the limit
$L \rightarrow \infty$. Far away from the point $x=0$, we will see that the
leading order oscillatory term in the density has a logarithmic prefactor
the interpretation of this will be provided in section 6.

In the ground state $\Psi_0^{(0)}$ of the noninteracting system, the $N$ 
particles occupy the lowest 
energy states described by (\ref{wf2}) with $n=1,2, \cdot \cdot \cdot , N$.
The perturbation in (\ref{pp1}) will connect this state to two types of states:

\noindent (A) a state of the type $\Psi^{(0)} (n^\prime ; n)$ which differs 
from the ground state $\Psi_0^{(0)}$ in that only one particle is excited from
the level $n$ to the level $n^\prime$, where $1 \le n \le N$ and $n^\prime >N$.

\noindent (B) a state of the type $\Psi^{(0)} (n_1^\prime , n_2^\prime ; n_1 ,
n_2)$ which differs from the ground state $\Psi_0^{(0)}$ in that only two 
particles are excited from the levels $n_1$ and $n_2$ to the levels 
$n_1^\prime$ and $n_2^\prime$, where $1 \le n_2 < n_1 \le N$ and $n_1^\prime >
n_2^\prime > N$.

The matrix elements of (\ref{pp1}) between $\Psi_0^{(0)}$ and states of
type A are given by
\bea
V_{n^\prime ; n} ~&\equiv & ~\langle \Psi^{(0)} (n^\prime ; n) \vert V \vert
\Psi_0^{(0)} \rangle ~=~ \frac{\pi^2}{8mcL^3} ~A_{n^\prime ; n} ~, \non \\
{\rm where} ~~ A_{n^\prime ,n} ~&=&~ 3 (n^2 ~+~ n^{\prime 2}) ~+~ 10 n 
n^\prime ~~ {\rm if} ~~ n^\prime + n ~~ {\rm is ~ even ~ and} ~~ 1 \le 
\frac{n^\prime + n}{2} \le N ~, \non \\
&=&~ -~ 3 (n^2 ~+~ n^{\prime 2}) ~+~ 10 n n^\prime ~~ {\rm if} ~~ n^\prime
+ n ~~ {\rm is ~ even ~ and} ~~ 1 \le \frac{n^\prime - n}{2} \le N ~. \non \\
&=&~ 0 ~~~ {\rm otherwise} ~.
\label{matel1}
\eea
 
The matrix elements of (\ref{pp1}) between $\Psi_0^{(0)}$ and states of type B
will be denoted by $V_{n_1^\prime , n_2^\prime; n_1 , n_2}$; however, we do
not need the expressions for these for reasons which will soon become clear.

Following (\ref{pert}), we see that the ground state wave function up to
order $1/c$ is given by
\bea
\Psi_0 = & & \Psi_0^{(0)} ~+~ \sum_{n^\prime >N} \sum_{1 \le n \le N} 
\Psi^{(0)} (n^\prime ; n) \frac{V_{n^\prime ; n}}{(n^2 - n^{\prime 2})\pi^2 /
(2mL^2)} \non \\
& & + ~\sum_{n_1^\prime > n_2^\prime > N} \sum_{1 \le n_2 < n_1 \le N} 
\Psi^{(0)} (n_1^\prime , n_2^\prime ; n_1 , n_2) \frac{V_{n_1^\prime , 
n_2^\prime ; n_1 , n_2}}{(n_1^2 + n_2^2 - n_1^{\prime 2} - n_2^{\prime 2})
\pi^2 / (2mL^2)} ~. \non \\
& &
\label{psi1}
\eea
We can now compute the density using (\ref{dens1}). To order $1/c$, we see that
we get a contribution from $\langle \Psi_0^{(0)} \vert \psi_0^{(0)} \rangle$ 
which is given in (\ref{dens2}), and a contribution from the cross-terms 
$\langle \Psi_0^{(0)} \vert \Psi^{(0)} (n^\prime ; n) \rangle$ arising
from the type A states in (\ref{psi1}). There is {\it no} contribution from the
cross-terms arising from the type B states since the integration over $dx_2 
\cdot \cdot \cdot dx_N$ ensures that $\langle \Psi_0^{(0)} \vert \Psi^{(0)} 
(n_1^\prime , n_2^\prime ; n_1 , n_2) \rangle$ does not contribute to the 
density. The vanishing of the contribution of type B states leads to a 
major simplification in the calculations. 

We now find that 
\beq
\rho (x) ~-~ \rho_0 ~+~ \frac{\sin (2 k_F x)}{2\pi x} ~=~ \frac{1}{c L^2} ~
\sum_{n^\prime , n} ~\sin (\frac{\pi n^\prime x}{L}) ~\sin (\frac{\pi nx}{L})~
\frac{A_{n^\prime ; n}}{n^2 - n^{\prime 2}} ~.
\label{dens3}
\eeq
where $A_{n^\prime ; n}$ is given in (\ref{matel1}). We now go to the 
thermodynamic limit. Let us introduce the variables $k = \pi n /L$ and 
$k^\prime = \pi n^\prime /L$, so that $0 \le k \le k_F$ and $k^\prime 
\ge k_F$. We replace the sums over $n$ and $n^\prime$ by $(L/\pi) 
\int dk$ and $(L/\pi) \int dk^\prime$ respectively. The condition for 
non-vanishing matrix elements in (\ref{matel1}), namely, that
$n^\prime + n$ must be an even integer, implies that we should put a factor
of $1/2$ in front of the integrals over $k$ and $k^\prime$. Further, 
the conditions that $(n^\prime \pm n)/2$ must be an integer lying in the range
$1$ to $N$ turn into the conditions $0 \le k^\prime \pm k \le 2 k_F$.
After some manipulations, we find that (\ref{dens3}) can be written in the form
\bea
& & \rho (x) ~-~ \rho_0 ~+~ \frac{\sin (2 k_F x)}{2\pi x} \non \\
& & = - ~\frac{1}{2 \pi^2 c} ~\int_{-k_F}^{k_F} dk~ \int_{\pi 
\rho_0}^{2 k_F -k} dk^\prime ~\frac{\sin (kx) \sin 
(k^\prime x)}{k^{\prime 2} - k^2} ~[~ 3 (k^{\prime 2} + k^2) +
10 k^\prime k ~] ~.
\label{dens4}
\eea
We now define two new variables $u=(k^\prime + k)/(2 k_F)$ and 
$v=(k^\prime - k)/(2 k_F)$. In terms of these variables, the limits of 
the integrals over $k$ and $k^\prime$ in (\ref{dens4}) are equivalent either 
to $\int_0^1 du \int_{1-u}^{1+u} dv$ or to $\int_0^2 dv \int_{|v-1|}^1 du$.
Finally, we can write $\sin (kx) \sin (k^\prime x) = (1/2) [\cos (2 k_F x v) 
- \cos (2 k_F x u)]$. We can now do one of the integrals, either 
over $u$ or over $v$, to obtain
\bea
& & \rho (x) ~-~ \rho_0 ~+~ \frac{\sin (2 k_F x)}{2\pi x} \non \\
& & = ~- ~\rho_0 g ~[~ \int_0^2 ~dv ~\cos (2 k_F x v) ~\{ 2 - v + 
\frac{v}{2} ~{\rm ln} \vert v -1 \vert \} \non \\
& & ~~~~~~~~~~~~~~~~ + ~\int_0^1 ~du ~\cos (2 k_F x u) ~\{ 1 - 2u ~{\rm
ln} (\frac{1+u}{1-u} ) \} ~] ~.
\label{dens5}
\eea
As shown in the appendix, the total hole in the density is found to be 
\beq
\int_0^\infty ~[~ \rho (x) - \rho_0 ~]~ =~ - \frac{1}{4} ~-~ \frac{3g}{4} ~.
\label{hole}
\eeq

{}From (\ref{dens5}), we can obtain the asymptotic form of the density at 
large values of $x$ (i.e., for $|k_F x| >> 1$) as shown in the appendix. The 
density turns out to have an expansion in powers of $1/x$ multiplied by 
sines, cosines, and logarithms. The leading order terms are given by 
\beq
\rho (x) ~=~ \rho_0 ~-~ \frac{\sin (2 k_F x)}{2 \pi x} ~[~ 1 - 2g ~{\rm
ln} (4 \pi e^{C-1/2} \rho_0 x) ~]~ + ~3g ~\frac{\cos (2 k_F x)}{4x} ~,
\label{asymp1}
\eeq
where $C=0.5772 \ldots$ is Euler's constant. We will see in 
section 6 that the logarithm is a sign that the power of $x$ in the 
denominator of $\sin (2 k_F x)$ should really be $1+2g$ (plus terms of 
higher order in $g$), rather than 1. Namely, 
\beq
\frac{1}{(\rho_0 x)^{1+2g}} ~=~ \frac{1}{\rho_0 x} [1 ~-~ 2g ~{\rm ln} (\rho_0
x ) ]
\eeq
plus terms of higher order in $g$.

\vskip .5 true cm
\noindent {\bf 4. The \df Bose gas with a discontinuity in the interaction 
strength}
\vskip .3 true cm

In this section, we will consider another application of the pseudopotential
approach. We will consider a \df Bose gas on a circle, with the interaction 
parameter in (\ref{ham}) being equal to a large value $c$ for $0 < x < L/2$ 
and $\infty$ for $L/2 < x < L$. It is clear from (\ref{mu}) that the chemical 
potential will then be different in the two halves of the system; this would 
imply that a state which has the same density everywhere can no longer be close
to the ground state. We will compensate for this imbalance 
by adding a one-body potential in the region $0 < x < L/2$ which is equal to
\beq
\delta V ~=~ \frac{4 \pi^2 \rho_0^3}{3mc} ~.
\label{dv}
\eeq
This ensures that the chemical potential and the density are the same in the
two halves of the system; therefore the ground states of the noninteracting 
($c=\infty$) and interacting (large $c$) systems are smoothly connected to 
each other. We will therefore work with the following perturbation to the 
noninteracting system, 
\bea
V ~&=&~ -~ \frac{1}{mc} ~\sum_{1 \le i<j \le N} ~\delta^{\prime \prime} ( ~
x_i ~ -~ x_j ~)~ f(x_i) ~+~ \frac{4 \pi^2 \rho_0^3}{3mc} ~\sum_{1 \le i \le N}
f(x_i) ~, \non \\
{\rm where} ~~ f(x) ~&=&~ 1 ~~ {\rm for} ~~ 0 < x < L/2 ~, \non \\
&=&~ 0 ~~ {\rm for} ~~ L/2 < x < L ~.
\label{pp2}
\eea
We will then find that there is a reflection from the point of discontinuity
at $x=0$.

We will now compute the density of the system to first order in $1/c$. In the 
ground state $\Psi_0^{(0)}$ of the noninteracting system, the particles
occupy the levels given in(\ref{wf1}) with $n=0, \pm 1, \pm 2, \cdot \cdot 
\cdot , \pm (N-1)/2$ (assuming that $N$ is odd). As in 
section 3, the perturbation in (\ref{pp2}) connects this state 
$\Psi_0^{(0)}$ to states of types A and B. Type A states differ from 
$\Psi_0^{(0)}$ in that only one particle is excited from a level $n$ lying 
in the range $[-(N-1)/2, (N-1)/2]$ to a level $n^\prime$
lying outside that range, while type B states differ from $\Psi_0^{(0)}$ in 
that only two particle are excited from levels $n_1$ and $n_2$ lying in 
the range $[-(N-1)/2, (N-1)/2]$ to levels $n_1^\prime$ and $n_2^\prime$
lying outside that range. After using (\ref{pert}) to write the perturbed
ground state wave function to order $1/c$, we again find that the type
B states do not contribute to the density. 

The matrix element of (\ref{pp2}) between $\Psi_0^{(0)}$ and a type A state 
$\Psi^{(0)} (n^\prime ; n)$ is given by
\bea
& & V_{n^\prime ; n} ~\equiv ~\langle \Psi^{(0)} (n^\prime ; n) \vert V \vert
\Psi_0^{(0)} \rangle \non \\
& & =~ \frac{i4 k_F}{mcL^2 (n^\prime -n)} ~[~ n^\prime n ~ +~ 
\frac{N^2 -1}{12} ~-~ \frac{\rho_0^2 L^2}{3} ~] ~~ {\rm if} ~~ n^\prime - n ~~
{\rm is ~ odd} ~, \non \\ 
& & =~ 0 ~~ {\rm otherwise} ~.
\label{matel2}
\eea
We now go to the thermodynamic limit, and introduce the variables $k^\prime
=2\pi n^\prime /L$ and $k=2\pi n/L$. Thus $k$ lies in the range $[- k_F, k_F]$
while $k^\prime$ lies outside it. We replace the sums over 
$n^\prime$ and $n$ by the integrals $(L/2\pi ) \int dk^\prime$ and $(L/2\pi )
\int dk$ respectively, and put a factor of $1/2$ in front of the integrals to 
take care of the restriction in (\ref{matel2}) that $n^\prime - n$ must be an 
odd integer. We then find that the density is given by 
\beq
\rho (x) ~=~ \rho_0 ~+~ \frac{g}{\pi^2} ~\int_{- k_F}^{k_F} dk ~
\int_{|k^\prime | \ge k_F} ~dk^\prime ~\frac{\sin \{ (k^\prime -k)
x \}}{(k^{\prime 2} -k^2)(k^\prime -k)} ~[~ k^\prime k - k_F^2 ~] ~.
\eeq
{}From this expression, it is clear that the choice of the one-body potential
in (\ref{dv}) makes the matrix element for forward scattering ($k^\prime = k 
= k_F$ or $k^\prime = k = - k_F$) vanish. This will lead to the 
result below that $\rho (x) - \rho_0$ vanishes for $x \rightarrow \pm \infty$;
this is the reason for choosing the value of $\delta V$ as given in (\ref{dv}).

We now introduce the variables $u = (k^\prime + k)/(2k_F)$ and $v = 
(k^\prime - k)/(2 k_F)$. For $v \ge 0$, $u$ lies in the range $|v-1|$ to
$v+1$, while for $v \le 0$, $u$ lies in the range $v-1$ to $-|v+1|$. After
integrating over $u$, we obtain
\beq
\rho (x) ~=~ \rho_0 ~+~ \frac{g}{2\pi} ~\int_0^\infty ~dv ~\frac{\sin (2 k_F
xv)}{v^2} ~[~ 2v ~-~ (v^2 + 1) ~{\rm ln} (\frac{v+1}{|v-1|}) ~]~.
\label{dens7}
\eeq
{}From (\ref{dens7}), we can obtain the asymptotic behavior of the density as
$x \rightarrow \pm \infty$ using the methods given in the appendix. We find 
that
\bea
\rho (x) ~&=&~ \rho_0 ~-~ g ~\frac{\sin (2 k_F x)}{2\pi x} ~~{\rm for} ~~
x \rightarrow \infty ~, \non \\
&=&~ \rho_0 ~+~ g ~\frac{\sin (2 k_F x)}{2\pi x} ~~{\rm for} ~~
x \rightarrow - \infty ~.
\label{asymp2}
\eea

We will now see that the oscillatory terms in (\ref{asymp2}) can be interpreted
as being due to reflection of the particles from the point $x=0$. Consider a 
system in which fermions coming in from $x = \infty$ get reflected back from 
$x=0$ with an amplitude $r_+$. The density at large positive values of $x$ can
be shown to be of the form \cite{yue,lal}
\beq
\rho (x) ~=~ \rho_0 ~+~ \frac{i}{4\pi x} ~[~ r_+^\star ~e^{-i2k_F x} ~-~ r_+ ~
e^{i2k_F x} ~]~ .
\label{rplus}
\eeq
On comparing this with (\ref{asymp2}), we see that 
\beq
r_+ ~=~ -~ g ~.
\eeq
Similarly, if fermions coming in from $x = - \infty$ get reflected back at 
$x=0$ with an amplitude $r_-$, the density at large negative values of $x$ 
can be shown to be 
\beq
\rho (x) ~=~ \rho_0 ~+~ \frac{i}{4\pi x} ~[~ r_- ~e^{-i2k_F x} ~-~ r_-^\star ~
e^{i2k_F x} ~]~ .
\label{rminus}
\eeq
On comparing this with (\ref{asymp2}), we see that 
\beq
r_- ~=~ g ~.
\eeq
In section 6, we will see that the values of $r_+$ and $r_-$ obtained
here agree exactly with those obtained from bosonization.

\vskip .5 true cm
\noindent {\bf 5. The \df Bose gas with a non-zero pseudopotential in a finite
region}
\vskip .3 true cm

We will now extend the analysis of the previous section to the case where
the interaction parameter $c$ is non-zero over a finite length $l$, where
$l$ will be held fixed as we go to the thermodynamic limit $L \rightarrow
\infty$. Now there is no particular reason to choose the strength of the
one-body potential $\delta V$ in the finite region $0 < x < l$ to be the same 
as in (\ref{dv}); as long as $\delta V$ is small, the ground state of this 
system is smoothly connected to that of the noninteracting system
in which $1/c$ and $\delta V$ are zero everywhere. At the end of our analysis,
we will discover that the reflection from the finite region is zero, i.e., 
there is a resonance in the transmission, if either $l$ has certain special 
values (given by $\pi n/k_F$), or if $\delta V$ has a particular value.

We will work with the following perturbation to the noninteracting ($c=\infty$)
system, 
\bea
V ~&=&~ -~ \frac{1}{mc} ~\sum_{1 \le i<j \le N} ~\delta^{\prime \prime} ( ~
x_i ~ -~ x_j ~)~ f(x_i) ~+~ \delta V ~\sum_{1 \le i \le N} f(x_i) ~, \non \\
{\rm where} ~~ f(x) ~&=&~ 1 ~~ {\rm for} ~~ 0 < x < c ~, \non \\
&=&~ 0 ~~ {\rm for} ~~ c < x < L ~.
\label{pp3}
\eea
As in section 4, we will compute the density of the system to first order in 
$1/c$. Once again, we find that the perturbation in (\ref{pp3}) connects the 
ground state $\Psi_0^{(0)}$ of the noninteracting system to states of types A 
and B, and the type B states do not contribute to the density to order $1/c$.

The matrix element of (\ref{pp3}) between $\Psi_0^{(0)}$ and a type A state 
$\Psi^{(0)} (n^\prime ; n)$ is given by
\bea
& & \langle \Psi^{(0)} (n^\prime ; n) \vert V \vert \Psi_0^{(0)} \rangle 
\non \\
& & ~=~ \frac{i4 k_F}{L^2 (n^\prime -n)} ~[~ \frac{1}{mc} (n^\prime n ~+~ 
\frac{N^2 - 1}{12}) ~-~ \frac{L^2 \delta V}{4\pi^2 \rho_0} ~] ~[~ \exp 
\{ i2\pi (n - n^\prime) \frac{l}{L} \} ~-~ 1 ~] ~. \non \\
& &
\eea
We introduce the variables $k =2\pi n/L$ lying in the range $[- k_F, k_F]$ and
and $k^\prime =2\pi n^\prime /L$ lying outside that range. 
After going to the thermodynamic limit, we find that the density is given by
\bea
\rho (x) ~=~ \rho_0 ~-~\frac{g}{\pi^2} ~\int_{-k_F}^{k_F} dk ~\int_{|k^\prime |
\ge k_F} ~dk^\prime ~ & & [~ \frac{\sin \{ (k^\prime -k) (x-l)\} - \sin \{ 
(k^\prime - k)x \}}{(k^{\prime 2} -k^2)(k^\prime -k)} ~] \non \\
& & \times ~[~ k^\prime k + \frac{\pi^2 \rho_0^2}{3} - \frac{mc \delta 
V}{\rho_0} ~] ~.
\label{dens8}
\eea
After introducing the variables $u = (k^\prime + k)/(2 k_F)$ and $v = 
(k^\prime - k)/(2 k_F)$, and integrating over $u$, we obtain
\bea
\rho (x) ~=~ \rho_0 ~-~ \frac{g}{2\pi} ~\int_0^\infty ~dv & & [~ \frac{\sin \{
(2k_F (x-l)v \} - \sin \{ 2 k_F xv \}}{v^2} ~] \non \\
& & \times ~[~ 2v ~-~ (v^2 - \frac{1}{3} + \frac{mc\delta V}{\pi^2 \rho_0^3} )~
{\rm ln} (\frac{v+ 1}{|v-1|}) ~]~.
\label{dens9}
\eea
Using the methods given in the appendix, we find that the asymptotic 
behavior of the density as $x \rightarrow \pm \infty$ is given by
\bea
\rho (x) ~&=&~ \rho_0 ~+~ (\frac{2\rho_0}{3c} + \frac{m\delta V}{\pi^2 
\rho_0^2})~ \frac{\sin \{ 2 k_F (x-l)\} - \sin \{ 2 k_F x \}}{4 \pi x} ~~
{\rm for} ~~ x \rightarrow \infty ~, \non \\
&=&~ \rho_0 ~-~ (\frac{2\rho_0}{3c} + \frac{m\delta V}{\pi^2 \rho_0^2})~ 
\frac{\sin \{ 2 k_F (x-l)\} - \sin \{ 2 k_F x \}}{4 \pi x} ~~{\rm 
for} ~~ x \rightarrow - \infty ~.
\label{asymp3}
\eea

As in section 4, the oscillatory terms in (\ref{asymp3}) can be interpreted as
being due to reflection of the particles from the region $0 < x < l$. On 
comparing (\ref{asymp3}) with the expressions in (\ref{rplus}) and 
(\ref{rminus}), we find that
\beq
r_+ ~=~ - i ~(~ \frac{2\rho_0}{3c} ~+~ \frac{m\delta V}{\pi^2 \rho_0^2} ~) ~
\sin (k_F l) ~e^{-i k_F l} ~,
\eeq
and
\beq
r_- ~=~ - i ~(~ \frac{2\rho_0}{3c} ~+~ \frac{m\delta V}{\pi^2 \rho_0^2} ~) ~
\sin (k_F l) ~e^{i k_F l} ~.
\eeq
We observe that the reflection amplitudes are zero (and the transmission
probabilities are therefore equal to 1) if either
\beq
l ~=~ \frac{\pi n}{k_F} ~,
\label{res1}
\eeq
where $n$ is an integer, or 
\beq
\delta V ~=~ - ~\frac{2\pi^2 \rho_0^3}{3mc} ~.
\label{res2}
\eeq
We see that up to order $1/c$, the resonance condition in (\ref{res1}) is the 
same as for noninteracting fermions ($c=\infty$); equation (\ref{res1})
is familiar from scattering theory in one-particle quantum mechanics.
The condition in (\ref{res2}) appears to be new; note that if (\ref{res2}) 
is satisfied, then $r_{\pm}$ is zero for all values of the length $l$. From 
(\ref{dens8}), we see that the resonance conditions in 
(\ref{res1}-\ref{res2}) imply that the matrix element for backward scattering
($k^\prime = - k = k_F$ or $k^\prime = - k = - k_F$) is zero.

\vskip .5 true cm
\noindent {\bf 6. The Tomonaga-Luttinger liquid approach to the \df Bose gas}
\vskip .3 true cm

We will now discuss how the results obtained in sections 3 and 4 may be 
understood using the TLL description of the long wavelength properties of
a \df Bose gas. As mentioned in section 1, a TLL is described by two 
parameters $v$ and $K$. These parameters can be deduced if one knows the
energy of the low-lying states \cite{hald}. Let us assume that the model
is defined on a circle of length $L$ with periodic boundary conditions.
Let the energy of the ground state be $E_0 (N,L)$. Then one has the relation
\beq
\Bigl( ~\frac{\partial^2 E_0}{\partial N^2} ~\Bigr)_L ~=~ \frac{\pi v}{LK} ~.
\label{vk1}
\eeq
Next, for a model which is invariant under Galilean transformations (as is
true for the \df Bose gas), it turns out that 
\beq
vK = v_F ~.
\label{vf}
\eeq
Since the ground state energy of the \df Bose gas is exactly known \cite{lieb},
one can use (\ref{vk1}-\ref{vf}) to find the values of $v$ and $K$ for any 
value of the interaction parameter $c$. From (\ref{e0}), we find that 
\bea
K ~&=&~ 1 + 2g ~, \non \\
v ~&=&~ v_F ~(1 ~-~ 2g) ~,
\label{vk2}
\eea
up to order $g$.
 
The low-energy and long wavelength properties of a one-dimensional fermionic 
system are determined by the modes near the two Fermi points with momenta 
$\pm k_F$. The second-quantized fermion fields $\Psi$ can be written in terms 
of the the fields near the two Fermi points as
\beq
\Psi (x) ~=~ \Psi_R (x) ~e^{ik_F x} ~+~ \Psi_L (x) ~e^{-ik_F x} ~,
\label{psirl}
\eeq
where the subscripts $R$ and $L$ denote right-moving and left-moving 
respectively. Let us define the density operators for these two fields as
${\hat \rho}_R = \Psi_R^\dagger \Psi_R$ and ${\hat \rho}_L = \Psi_L^\dagger 
\Psi_L$. Consider a contact interaction of the form \cite{gogolin,hald}
\beq
V_{int} ~=~ \int dx ~[~ g_2 {\hat \rho}_R (x) {\hat \rho}_L (x) ~+~ 
\frac{g_4}{2}~ \{ {\hat \rho}_R^2 (x) ~+~ {\hat \rho}_L^2 (x) \} ~]~.
\label{int1}
\eeq
Then one can show that 
\bea
K ~&=&~ [ ~(~ v_F ~+~ \frac{g_4}{2\pi} ~-~ \frac{g_2}{2\pi} ~) ~/~ (~ v_F ~+~
\frac{g_4}{2\pi} ~+~ \frac{g_2}{2\pi} ~) ~]^{1/2} ~, \non \\
v ~&=&~ [ ~(~ v_F ~+~ \frac{g_4}{2\pi} ~-~ \frac{g_2}{2\pi} ~)~ (~ v_F ~+~
\frac{g_4}{2\pi} ~+~ \frac{g_2}{2\pi} ~) ~]^{1/2} ~.
\eea
Comparing these expressions with (\ref{vk2}), we see that for the \df Bose 
gas near the fermionic limit, 
\beq
\frac{g_2}{2\pi v_F} ~=~ \frac{g_4}{2\pi v_F} ~=~ - ~2g ~.
\label{g24}
\eeq

We can directly verify the expression for $g_2$ given in (\ref{g24}). Consider
a density-density interaction of the form 
\beq
V_{int} ~=~ \frac{1}{2} ~\int \int dx ~dy ~{\hat \rho} (x) ~U (x-y) ~
{\hat \rho} (y) ~,
\label{int2}
\eeq
where the density operator follows from (\ref{psirl}),
\bea
& & {\hat \rho} (x) ~=~ \Psi^\dagger (x) \Psi (x) \non \\
& & = \Psi_R^\dagger (x) \Psi_R (x) ~+~ \Psi_L^\dagger (x) \Psi_L (x) ~+~
\Psi_R^\dagger (x) \Psi_L (x) e^{-i2k_F x} ~+~ \Psi_L^\dagger (x) \Psi_R (x) 
e^{i2k_F x} ~. \non \\
& &
\label{densop}
\eea
If the two-body potential $U(x)$ has a short range, a comparison of 
(\ref{int1}) and (\ref{int2}) shows that the parameter $g_2$ 
is related to the Fourier transform of $U(x)$ as
\beq
g_2 ~=~ {\tilde U} (0) - {\tilde U} (2k_F ) ~.
\label{g2}
\eeq
Comparing (\ref{int2}) and (\ref{pp1}), we see that 
\beq
U (x) ~=~ - ~\frac{1}{mc} ~\delta^{\prime \prime} (x) ~.
\eeq
Then (\ref{g2}) implies that
\beq
g_2 ~=~ - ~\frac{4k_F^2}{mc} 
\eeq
which agrees with (\ref{g24}).

Now we consider the bosonized form of the TLL theory \cite{gogolin,hald}. The 
bosonic field $\phi (x,t)$ is governed by the Lagrangian density
\beq
{\cal L} ~=~ \frac{1}{2vK} ~\Bigl(~ \frac{\partial \phi}{\partial t} ~
\Bigr)^2 ~-~\frac{v}{2K} ~\Bigl(~ \frac{\partial \phi}{\partial x} ~\Bigr)^2 ~.
\label{lag}
\eeq
The equations of motion are given by $\partial^2 \phi /\partial t^2 = v^2
\partial^2 \phi /\partial x^2$. The excitations of the system therefore have 
the dispersion $\omega = v|k|$.

The technique of bosonization relates the fermi fields $\Psi_R$ 
and $\Psi_L$ to exponentials of the second-quantized boson field $\phi$. The 
exact relationship between the two fields depends on the geometrical situation.
Let us consider the model of section 3, where the system is defined on the 
half-line $x \ge 0$. We denote the incoming (left-moving) fermi field as 
$\Psi_L$ and the outgoing (right-moving) fermi field as $\Psi_R$. These are
not independent fields since one is related to the other by reflection at 
$x=0$. One can now "unfold" the half-line to the full line and define
all the fields to be purely left-moving \cite{gogolin}. The fermi fields 
${\tilde \Psi}_L$ on the full line are related to those on the half-line as 
\bea
{\tilde \Psi}_L (x) = \Psi_L (x) ~, \non \\
{\tilde \Psi}_L (-x) = \Psi_R (x) ~,
\label{psitil}
\eea
where $x > 0$. The bosonized form of this fermionic theory also contains only 
a left-moving boson field $\phi_L$ defined on the full line; the two fields 
are related as 
\beq
{\tilde \Psi}_L (x) ~\sim ~ \exp ~[i ~{\sqrt \frac{\pi}{K}} ~\{ \phi_L (x) + 
\phi_L (-x) \} ~+~ i ~{\sqrt {\pi K}} ~\{ \phi_L (x) - \phi_L (-x) \}] ~.
\label{boson}
\eeq 

Now we can compute the fermion density which is equal to an expectation value 
in the ground state, $\rho (x) = \langle \Psi^\dagger (x) \Psi (x) \rangle$, 
where $x>0$. Following (\ref{densop}) and (\ref{psitil}), this is given by
\bea
\rho (x) = & & \langle {\tilde \Psi}_L^\dagger (x) {\tilde \Psi}_L (x) \rangle
~+~ \langle {\tilde \Psi}_L^\dagger (-x) {\tilde \Psi}_L (-x) \rangle \non \\
& & +~\langle {\tilde \Psi}_L^\dagger (-x) {\tilde \Psi}_L (x) \rangle ~
e^{-i2k_F x} ~+~ \langle {\tilde \Psi}_L^\dagger (x) {\tilde \Psi}_L (-x) 
\rangle ~ e^{i2k_F x} ~.
\label{rhopsi}
\eea
On using the bosonization expression (\ref{boson}), we find that the first two
terms on the right hand side of (\ref{rhopsi}) are independent of $x$; they 
give rise to a constant which is $\rho_0$. The last two terms in (\ref{rhopsi})
give
\beq
\langle \exp ~[\pm i 2 {\sqrt {\pi K}} \{ \phi_L (-x) - \phi_L (x) \} ] 
\rangle ~\sim ~\frac{1}{x^K} ~.
\eeq
Including the factors of $e^{\pm i2k_F x}$ in (\ref{rhopsi}), we find that
\beq
\rho (x) ~-~ \rho_0 ~\sim ~ \frac{\sin (2k_F x)}{x^K} ~.
\eeq
We thus see that in the presence of a hard wall, the density of a TLL far
from the wall has an oscillatory piece whose amplitude decays as $1/x^K$.
If $K$ is close to 1 as in (\ref{vk2}), we see that the amplitude has
an expansion in powers of $g$ which is given by $(1 - 2g ~{\rm ln} x)/x$.
This is exactly what we found in section 3.

We now turn to the model of section 4, where the Luttinger parameters are given
by $(v,K)$ for $x>0$ and by $(v_F,1)$ for $x<0$. We can use the Lagrangian 
density in (\ref{lag}) with these parameters to find the equations of motion 
for the boson fields. The matching conditions at $x=0$ turn out to be 
\cite{safi} 
\bea
\phi (x=0- , t) ~&=&~ \phi (x=0+ , t) ~, \non \\
v_F ~\Bigl(~ \frac{\partial \phi (x,t)}{\partial x} ~\Bigr)_{x=0-} ~&=&~ 
\frac{v}{K} ~\Bigl(~ \frac{\partial \phi (x,t)}{\partial x} ~\Bigr)_{x=0+} ~.
\label{match}
\eea
We can now consider what happens when a wave is incident from $x=-\infty$. The
equations of motion give
\bea
\phi (x,t) ~&=&~ \exp ~[ik(x-v_Ft)] ~+~ r_- ~\exp ~[-ik(x+v_Ft)] ~~ {\rm 
for} ~~ x < 0 ~, \non \\
&=&~ t_- ~\exp ~[i \frac{kv_F}{v} ~(x-v t)] ~~ {\rm for} ~~ x > 0 ~.
\eea
The matching conditions in (\ref{match}) now lead to the following 
expressions for the reflection and transmission amplitudes
\bea
r_- ~&=&~ \frac{K-1}{K+1} ~, \non \\
t_- ~&=&~ \frac{2K}{K+1} ~.
\eea
Note that current conservation is satisfied since $v_F (1-r_-^2) = v t^2$ and
$v_F = vK$. Similarly, for a wave incident from $x=\infty$, we find that 
\bea
r_+ ~&=&~ \frac{1-K}{K+1} ~, \non \\
t_+ ~&=&~ \frac{2}{K+1} ~,
\eea
which satisfies current conservation since $v (1-r_+^2 ) = v_F t_+^2$.
Upon using (\ref{vk2}), we see that the reflection amplitudes $r_{\pm}$ 
obtained here agree with those obtained in section 4.

We therefore have the interesting result that a discontinuity in the
interaction parameters is sufficient to cause scattering, even if there
is no other scattering mechanism (like an impurity) present in the system.
The calculations presented in section 4 can be viewed as a microscopic
derivation of this interesting phenomenon which had earlier been obtained
only from bosonization \cite{safi}. 

The results in section 5 have implications for the subject of transport through
a quantum wire which is sometimes modeled as a TLL of finite length which is 
bounded on the two sides by Fermi liquid leads \cite{safi,maslov,kane}. In 
these models, therefore, the Luttinger parameters change discontinuously at 
the contacts between the quantum wire and its leads. For the 
case of two identical impurities in a TLL, it is known that the transmission 
resonances are infinitely sharp at zero temperature \cite{kane}. Although
the model we have studied in section 5 has two points of discontinuity in
the interaction parameter, rather than two impurities, it is possible that the
structure of the transmission resonance will be found to change significantly
if we go up to higher orders in the interaction parameter $1/c$.
 
\vskip .5 true cm
\noindent {\bf 6. Discussion}
\vskip .3 true cm

We have used the pseudopotential approach to study the behavior of a system 
of fermions with weak attractive interactions. The various situations we have 
considered are not solvable by the Bethe ansatz \cite{lieb}. This is 
because the Bethe ansatz only works in models which are both invariant under 
translations and have $N$ commuting operators including the Hamiltonian. In 
such systems, the momenta of the $N$ particles $k_1, k_2, \cdot \cdot \cdot , 
k_N$ are good quantum numbers, and the wave function can be found exactly as 
a superposition of $N!$ plane waves. In the absence of translation invariance,
there are reflections (at a hard wall or at a point of discontinuity in the 
interaction strength) due to which the particle momenta are no longer good 
quantum numbers. The wave functions are then no longer a superposition of a 
finite number of plane waves, and they cannot be found exactly.

When the Bethe ansatz fails, the pseudopotential approach seems
to be the only way to obtain exact results near the fermionic limit, although
calculational difficulties may restrict its use to low orders in perturbation 
theory. We have shown how exact expressions for the density can be obtained
in certain situations. While the agreement between our results and those 
obtained from bosonization for the asymptotic behavior of the density is 
satisfying, we should also emphasize that there are relatively few models 
with interactions in which something can be computed at all distances. 

Our methods can be applied to other problems involving weakly interacting 
fermions in one dimension. For instance, one can study the Kane-Fisher model 
of a single impurity placed in a TLL \cite{kane}, and the effect of a 
junction of three or more semi-infinite wires \cite{lal}. While these problems
have been studied earlier using bosonization (and other methods which are
only valid at long wavelengths \cite{yue}), it may be interesting to apply the
pseudopotential method to these situations since we may be able to obtain 
expressions for certain quantities which are valid at all distances. Finally, 
one can use the pseudopotential method to study dynamical quantities like the
conductance of a finite length TLL at finite frequencies.

After this work was completed, we found a paper which discusses some properties
of the one-dimensional Bose-Hubbard model at low densities \cite{caza}; the
continuum Hamiltonian which governs that system is essentially the same as the
one studied by us.

\vskip .5 true cm
\noindent {\bf Acknowledgments}
\vskip .3 true cm

I thank Siddhartha Lal and Sumathi Rao for numerous discussions about 
bosonization and quantum wires. I acknowledge financial support from a Homi 
Bhabha Fellowship and the Council of Scientific and Industrial Research, 
India through grant no. 03(0911)/00/EMR-II.

\vskip .5 true cm
\noindent {\bf Appendix}
\vskip .3 true cm

In this appendix, we will discuss some methods for computing the various 
expressions for the density presented above.

Let us start with the expression for the hole in (\ref{hole}). Suppose
that we have a function $f(x)$ defined by the integral
\beq
f(x) ~=~ \int_0^{u_0} ~du ~\cos (axu) ~h(u) ~,
\eeq
where $h(u)$ is finite and continuous at $u=0$. Then we find 
\bea 
\int_0^\infty ~dx ~f(x) ~&=&~ {\rm lim}_{\alpha \rightarrow 0} \int_0^{u_0} ~
du ~h(u) ~\int_0^\infty ~dx ~\cos (axu) ~e^{-\alpha x^2} ~, \non \\
&=&~ {\rm lim}_{\alpha \rightarrow 0} ~\frac{1}{2} ~
{\sqrt {\frac{\pi}{\alpha}}} ~\int_0^{u_0} ~du ~h(u) ~e^{-a^2 u^2 /(4 \alpha)},
\non \\
&=&~ \frac{\pi}{2a} ~h(0) ~.
\eea
On applying this to (\ref{dens5}), we obtain (\ref{hole}).

Next, we discuss how to obtain asymptotic expressions (for $ax >> 1$)
for functions of the type 
\bea
f(x) ~&=&~ \int_0^{u_0} ~du ~\cos (axu) ~h(u) ~, \non \\
g(x) ~&=&~ \int_0^{u_0} ~du ~\sin (axu) ~h(u) ~.
\label{fgx}
\eea
If $h(u)$ is finite and continuous for all values of $u$ in the range 
$[0,u_0]$, then the leading order expressions for (\ref{fgx}) are of order 
$1/x$, and they are obtained by integrating the functions $\cos (axu)$ and 
$\sin (axu)$. Namely, 
\bea
f(x) ~&=&~ \frac{1}{ax} ~[~ h(u_0) ~\sin (axu_0 ) ~] ~, \non \\
g(x) ~&=&~ \frac{1}{ax} ~[~ h(0) ~-~ h(u_0) ~\cos (axu_0 ) ~] ~,
\label{asymp4}
\eea
plus terms of order $1/x^2$ and higher. These formulae are valid even if
$u_0 = \infty$, provided that $h (\infty) =0$.

Now suppose that the function $h(u)$ has a logarithmic divergence at one
point. Some examples of this are the integrals \cite{grad} 
\bea
\int_0^1 ~du ~\sin (axu) ~{\rm ln} u ~&=&~ - ~\frac{1}{ax} ~[~ {\rm ln} (ax) 
+ ~C ~] ~, \non \\
\int_0^1 ~du ~\cos (axu) ~{\rm ln} u ~&=&~ - ~\frac{\pi}{2ax} ~,
\label{lnint1}
\eea
plus terms of order $1/x^2$ and higher. (Here $C=0.5722 \ldots$ is 
Euler's constant). From (\ref{lnint1}), one can show that
\bea
\int_0^1 ~du ~\sin (axu) ~{\rm ln} (1-u) ~&=&~ ~\frac{1}{ax} ~[~ {\rm ln} 
(ax) \cos (ax) ~+~ C \cos (ax) ~-~ \frac{\pi}{2} \sin (ax) ~]~ , \non \\
\int_0^1 ~du ~\cos (axu) ~{\rm ln} (1-u) ~&=&~ - ~\frac{1}{ax} ~[~ {\rm ln} 
(ax) \sin (ax) ~+~ C \sin (ax) ~+~ \frac{\pi}{2} \cos (ax) ~]~ , \non \\
\label{lnint2}
\eea
plus terms of order $1/x^2$ and higher.

Now consider a function $h(u)$ in (\ref{fgx}) which contains a logarithmic 
divergence at $u=0$ of the form $b ~{\rm ln} u$, where $b$ is some constant. 
Then we define a new function ${\tilde h} (u) = h(u) - b ~{\rm ln} u$ which is
finite for all values of $u$ in the range of integration. We can then integrate
over ${\tilde h} (u)$ using (\ref{fgx}-\ref{asymp4}), and over $b ~{\rm ln} u$
using (\ref{lnint1}). Combining the two gives the result of integrating over
$h(u)$. Similarly, we can compute the integrals in (\ref{fgx}) if 
$h(u)$ has a logarithmic divergence at $u=u_0$. Finally, we can also compute 
integrals over functions $h(u)$ in (\ref{fgx}) which have a logarithmic 
divergence at a point $u_1$ which lies inside the range $[0,u_0]$. We simply 
divide the integrals in (\ref{fgx}) into two parts, one over the range 
$[0,u_1]$ and the other over the range $[u_1,u_0]$. Then each of these 
integrals can be computed as explained above. Using all these methods, one 
can derive the expressions in (\ref{asymp1}) and (\ref{asymp2}).

\end{document}